\pdfoutput=1
\documentclass{PoS}
\usepackage[]{gensymb}
\usepackage{enumerate}
\newcommand{\eh}[1]{\,\mathrm{#1}}
\newcommand{\gev}{\eh{GeV}}

\newcommand{\tev}{\eh{TeV}}

\newcommand{\dg}{^{\circ}}

\renewcommand{\epsilon}{\varepsilon}

\newcommand{\hess}{H.E.S.S.}
\newcommand{\nectar}{NECTAr}
\newcommand{\hessI}{H.E.S.S.~I}
\newcommand{\hessIu}{H.E.S.S.~I Upgrade}

\newcommand{\zmq}{{\O}MQ}

\newcommand{\fref}[1]{Fig.~\ref{#1}}

\title{Hardware and software architecture of the upgraded H.E.S.S. cameras}

\ShortTitle{\hess\ upgrade hardware and software}

\author{\speaker{S.~Klepser} $^a$,
T.~Ashton$^b$,
M.~Backes$^h$,
A.~Balzer$^c$,
D.~Berge$^c$,
S.~Bonnefoy$^a$
F.~Brun$^d$,
T.~Chaminade$^d$,
E.~Delagnes$^d$,
G.~Fontaine$^f$,
M.~F\"u{\ss}ling$^a$,
G.~Giavitto$^a$,
B.~Giebels$^f$,
J.F.~Glicenstein$^d$,
T.~Gr\"aber$^a$,
J.A.~Hinton$^{b,g}$,
A.~Jahnke$^i$,
M.~Kossatz$^a$,
A.~Kretzschmann$^a$,
V.~Lefranc$^{a,d}$,
H.~Leich$^a$,
J.P.~Lenain$^e$,
H.~L\"udecke$^a$,
I.~Lypova$^a$,
P.~Manigot$^f$,
V.~Marandon$^g$,
E.~Moulin$^d$,
M.~de~Naurois$^f$,
P.~Nayman$^e$,
S.~Ohm$^a$,
M.~Penno$^a$,
D.~Ross$^b$,
D.~Salek$^c$,
M.~Schade$^a$,
T.~Schwab$^g$,
K.~Shiningayamwe$^h$
C.~Stegmann$^a$,
C.~Steppa$^a$,
J.~Thornhill$^b$,
F.~Toussenel$^e$\\
\llap{$^a$}DESY, D-15738 Zeuthen, Germany \\
\llap{$^b$}Department of Physics and Astronomy, The University of Leicester, University Road, Leicester, LE1 7RH, United Kingdom \\
\llap{$^c$}GRAPPA, Anton Pannekoek Institute for Astronomy, University of Amsterdam,  Science Park 904, 1098 XH Amsterdam, The Netherlands \\
\llap{$^d$}DSM/Irfu, CEA Saclay, F-91191 Gif-Sur-Yvette Cedex, France \\
\llap{$^e$}Sorbonne Universit\'es, UPMC Universit\'e Paris 06, Universit\'e Paris Diderot, Sorbonne Paris Cit\'e, CNRS, Laboratoire de Physique Nucl\'eaire et de Hautes Energies (LPNHE), 4 place Jussieu, F-75252, Paris Cedex 5, France \\
\llap{$^f$}Laboratoire Leprince-Ringuet, Ecole Polytechnique, CNRS/IN2P3, F-91128 Palaiseau, France \\
\llap{$^g$}Max-Planck-Institut f\"ur Kernphysik, P.O. Box 103980, D 69029 Heidelberg, Germany \\
\llap{$^h$}University of Namibia, Department of Physics, Private Bag 13301, Windhoek, Namibia\\
\llap{$^i$}JA consulting, St Michael Park 23, Avis, Windhoek, Namibia\\
}

\abstract{
In 2015/16, the photomultiplier cameras of the H.E.S.S. Cherenkov telescopes CT1-4 have undergone a major upgrade. The entire electronics has been replaced, using NECTAr chips for the front-end readout. A new ventilation system has been installed and several auxiliary components have been replaced. Besides this, the internal control and readout software was rewritten from scratch in a modern and modular way. Ethernet technology was used wherever possible to ensure both flexibility, stability and high bandwidth. An overview of the installed components will be given.
}

\FullConference{35th International Cosmic Ray Conference --- ICRC2017\\
		10--20 July, 2017\\
		Bexco, Busan, Korea}

\begin{document}
\section{Introduction}
The H.E.S.S. experiment is located in the Khomas Highland of Namibia  \cite{hess_crab}.
From January 2004 to July 2012, it consisted of an array of four 12-meter Imaging Atmospheric Cherenkov Telescopes (IACTs) named CT1-4. They form a square with a side length of $120\eh{m}$, have an effective mirror area of $107\eh{m^2}$, detect cosmic gamma rays in the $100\gev$ to $100\tev$ energy range, and cover a field of view of $5\dg$ in diameter.
In July 2012, a fifth telescope (CT5) with a much larger mirror was commissioned. It provides a lower threshold of around $30\gev$ and a higher event rate of $\sim 1.5\eh{kHz}$. Operated in coincidence mode with the smaller CT1-4, this would have allowed also for a lower stereoscopic threshold of the entire array, but the readout dead time of the CT1-4 cameras \cite{oldcameras} was $\sim 450\eh{\mu s}$, which did not allow for a substantial lowering of the trigger threshold.

The primary goal of the upgrade of the CT1-4 cameras presented here was therefore to reduce the dead time of the cameras when operating with CT5. Another aspect was to reduce the downtime of the cameras: after a decade of operation in the Namibian Savannah, with its harsh environment, the failure rate had begun to increase and a redesign of the cameras with an enhanced filtered ventilation system had become desirable to stabilise operations.

The first new camera was installed in July 2015 in the CT1 telescope (see \fref{fig:overview}, left). After its extensive testing and validation, the other three cameras were produced and installed already in September 2016. Since February 2017, all four cameras are in routine operation and part of the \hess\ observation program.

The following will report about the deployed hardware and software of the new components. The performance is described in more detail in a second contribution to this conference \cite{hess1u_perf}.

\section{Hardware architecture}

\subsection{Electronic front end}

In the course of the camera upgrade, almost every component inside the camera was replaced, except for the photomultiplier tubes (PMTs), their high voltage bases and the calibration
units. A scheme of the logical blocks of the new cameras and their interplay can be seen in \fref{fig:architecture}.

\begin{figure}
  \centering
  \includegraphics[height=0.35\textwidth]{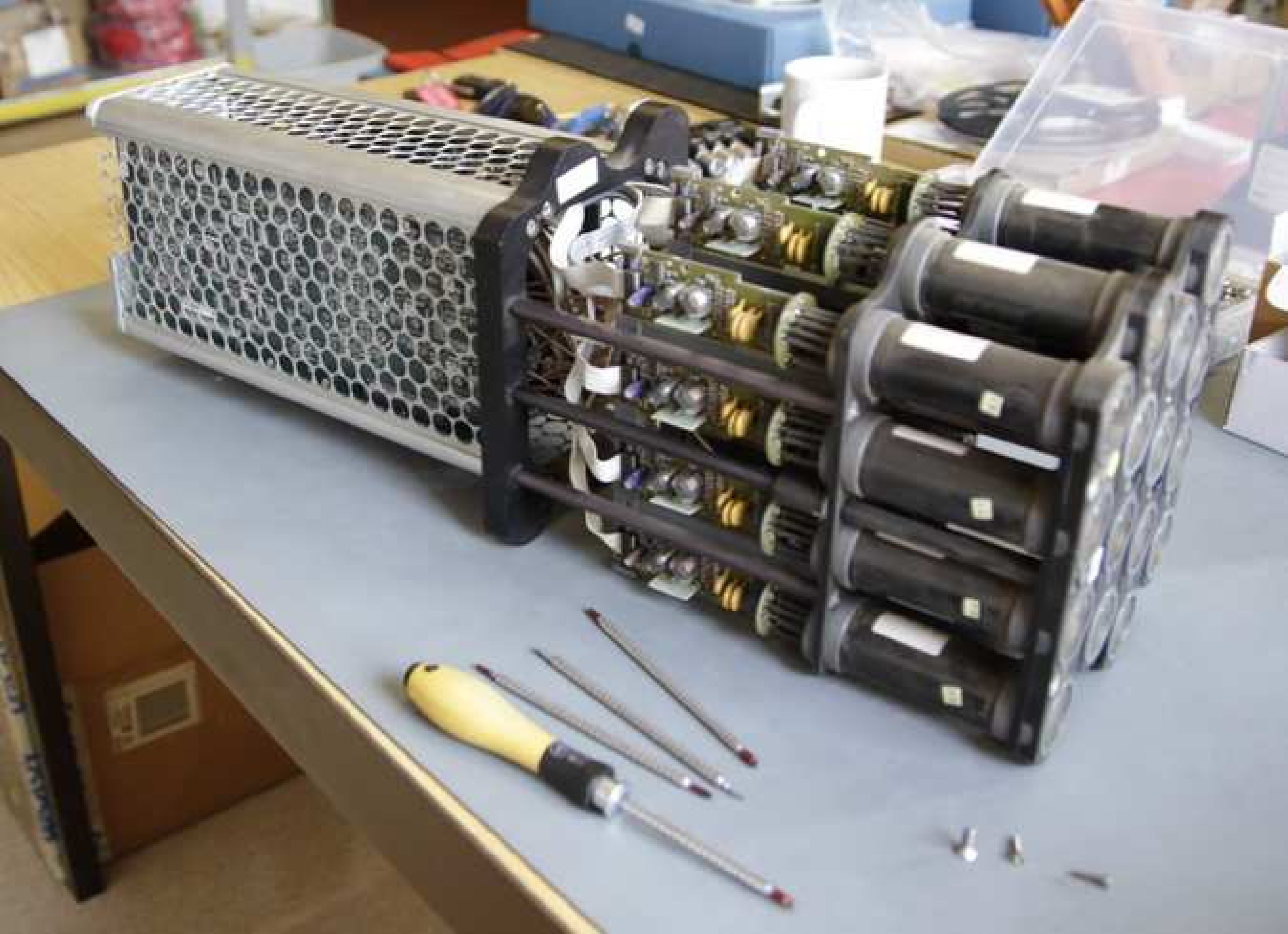}
  \includegraphics[height=0.35\textwidth]{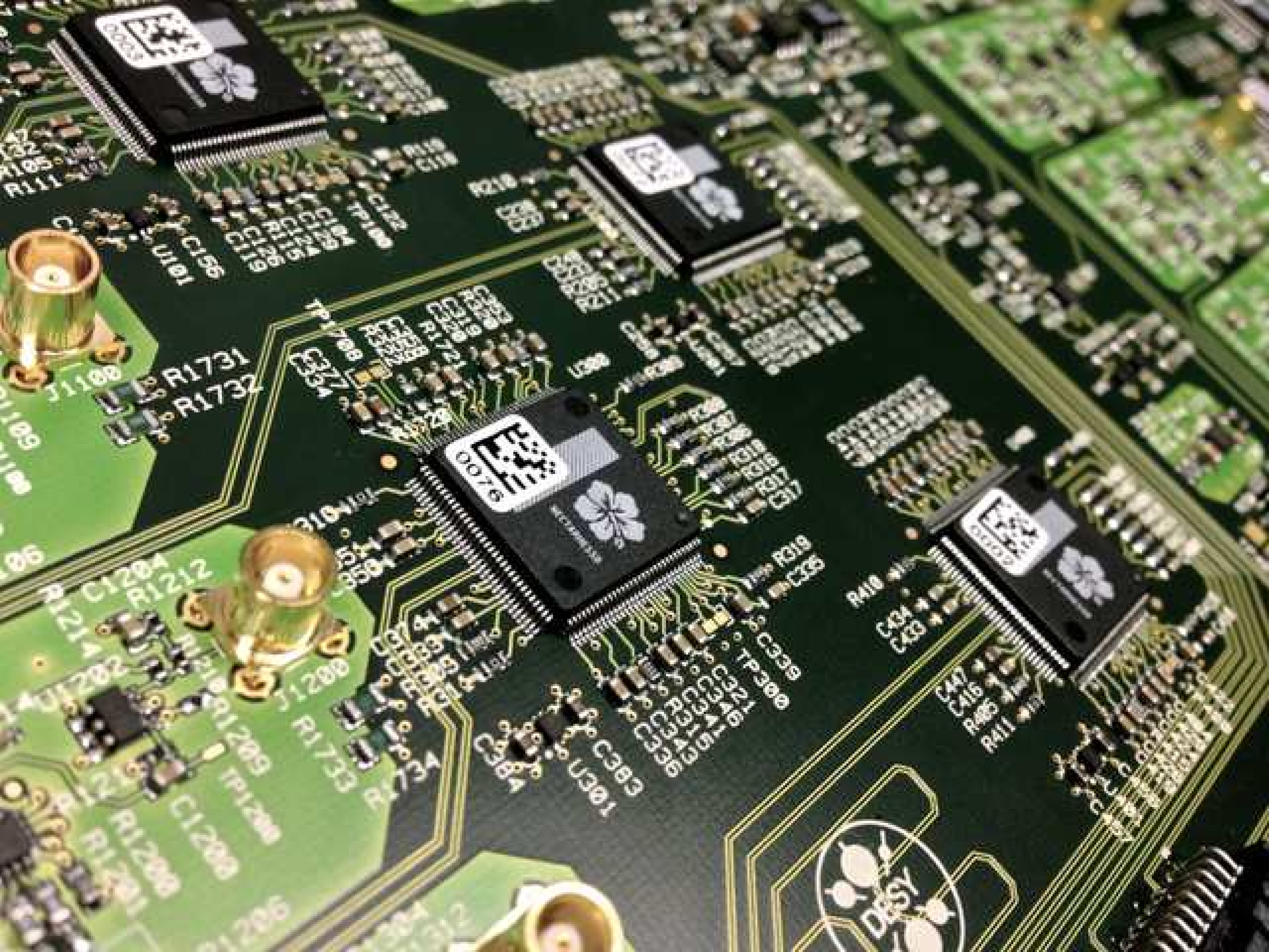}
  \caption{Left: One newly assembled \hess\ camera drawer. Right: Close-up view of the analogue trigger board and mounted \nectar.}
  \label{fig:detailed_views}
\end{figure}

The most important element of the upgraded cameras contributing to the
reduction of the dead time is the readout, built around the NECTAr analog
memory chip \cite{nectar}. It is located in the front-end part of the camera where Cherenkov light from particle showers
in the atmosphere is detected and digitized. The light sensors are 960 PMTs, sorted into modules called \emph{drawers} (\fref{fig:detailed_views}, left), which consist of 16 PMTs each and which are also responsible for the amplification and readout of the PMT signals.
A drawer consists of two
analogue boards (\fref{fig:detailed_views}, right), each hosting 8 analogue channels. They are mounted vertically on top of a slow control
board, which is located at the bottom of the drawer. Each slow control board is connected to the back end through a \emph{connection board}, which receives power, sends data, and communicates trigger and operation settings to/from the back end of the camera.

\begin{figure}
  \centering
  \includegraphics[height=0.35\textwidth]{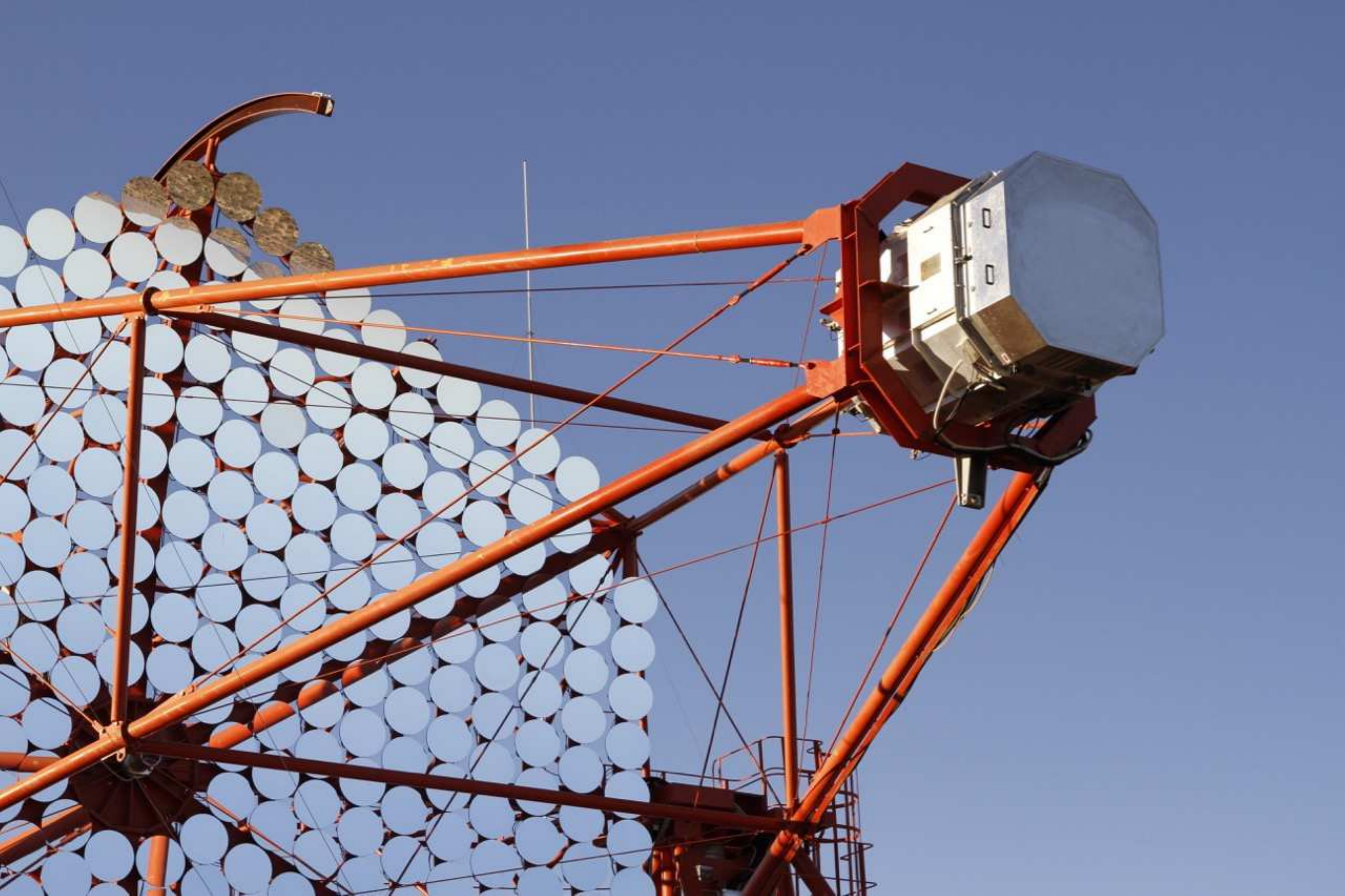}
  \includegraphics[height=0.35\textwidth]{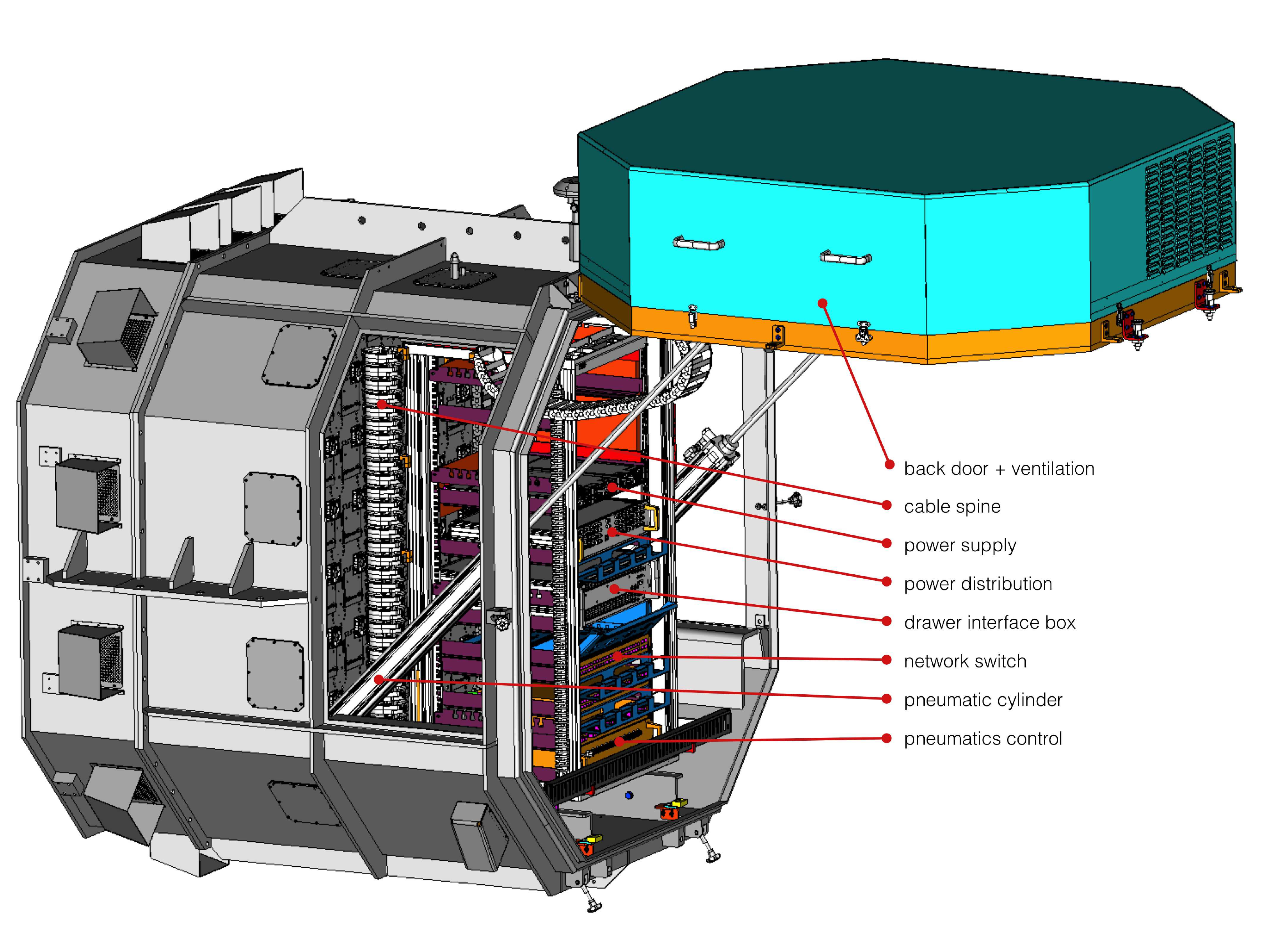}
  \caption{Left: The first \hessIu\ camera mounted on the CT1 telescope. Right: Technical sketch of the layout of the new cameras and the components of the back-end rack.}
  \label{fig:overview}
\end{figure}

\subsection{Electronic back end}

The back-end part of the camera consists of a so-called drawer interface box (DIB), a power supply unit, a power distribution box, an Ethernet switch and a pneumatics control unit (\fref{fig:overview}, right).

The DIB is the versatile heart of the new camera design. It combines several functionalities (see also \fref{fig:architecture}): 
\begin{itemize}
\item Trigger interface to 60 drawers and to the array central trigger
\item Clock distribution to the drawers
\item Camera (L1) trigger generation
\item Interface to various camera components, see \fref{fig:architecture}
\item Protection of PMT and camera electronics through a HV interlock and enable signal for the drawer power
\item GPS timestamping of events
\end{itemize}
With the technological advances of the past 20 years, the new back-end components could be designed and build in a much more compact way than in the original cameras. Instead of two racks, we now only needed one rack for all the units we built. There was even space to install a \emph{CTA auxiliary box} in some of the cameras. It is an interface to test components (flat-fielding unit, timing unit) that were prototyped for the Cherenkov Telescope Array (CTA).

\subsection{Mechanical upgrades}

Besides the reworked electronics, several mechanical aspects were also improved. A new ventilation concept was worked out that steadily injects filtered and
(optionally) heated air into the camera. It causes a steady overpressure in the
entire camera body, and leads to a constant air flow through the drawers and
all other small openings of the camera. This way a homogeneous cooling is
achieved, and dust is kept out of the camera. The ventilation was integrated in a new rear door of the cameras (see \fref{fig:overview}, right).

This rear door is heavier than the previous ones, so a pneumatic solution was implemented to open and close them. Along with this effort, also the pneumatic system of the entire camera was reworked and modernised.

\section{Network}

Most of the components designed for this upgrade are built such that they also constitute a device in an internal Ethernet network of the camera. This is achieved with an FPGA
(Cyclone IV) and an ARMv5-based $\mu$Computer Module (Stamp9G45) in these devices. The two communicate through a fast 100MBit/s memory bus. This solution allows for high flexibility and good control and communication with all subsystems. It improves the overall resilience of the camera.

\begin{figure}
  \centering
  \includegraphics[height=0.55\textwidth]{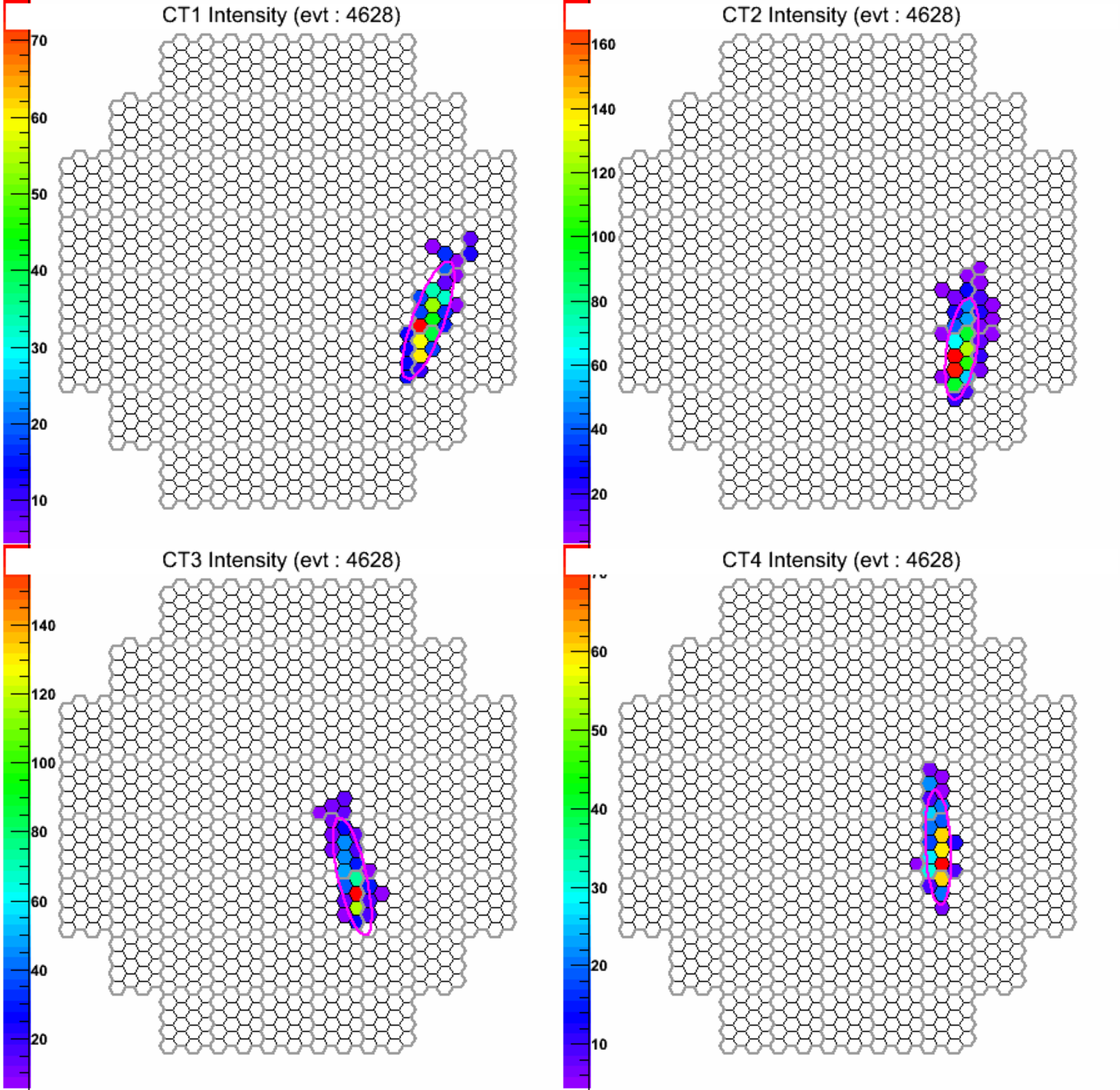}
    \includegraphics[height=0.55\textwidth]{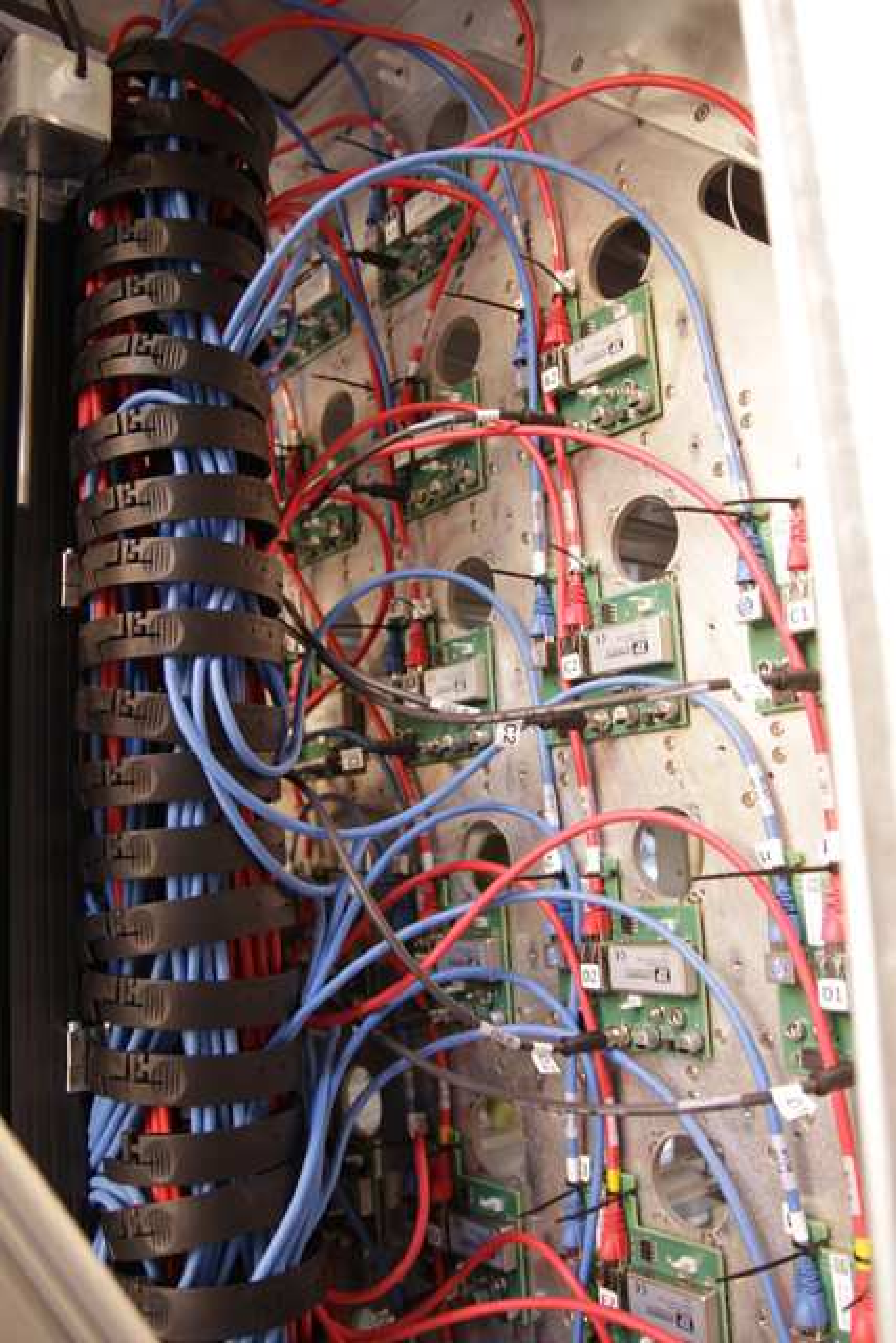}
  \caption{Left: Example four-telescope event recorded with the upgraded cameras of CT1-4. Right: Photograph of the cable spines and connection boards, used for the distribution of signal (red), trigger (blue) and power (black) cables.}
  \label{fig:event}
\end{figure}

The trigger and data  paths are realised using Cat.~6$_A$ and Cat.~6 Ethernet cables, respectively  (see  \fref{fig:event}, right). The power is transmitted through industry standard 4-wire
cables, ending with M8 connectors. The Ethernet and power cables are bundled in industrial cable spines, which were premounted in Europe and allowed for a fast cabling process on site.

\begin{figure}
  \centering
  \includegraphics[width=\textwidth]{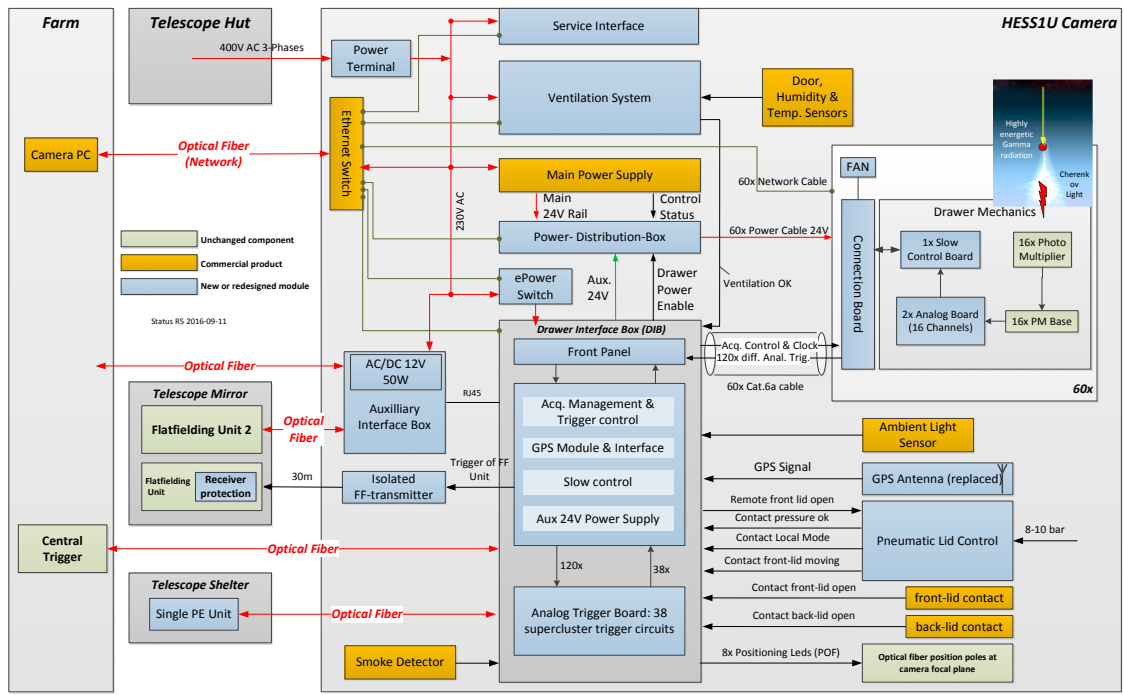}
  \caption{Schematic diagram illustrating the logical blocks of the \hessIu. The
  few  original \hessI\ systems are depicted as green boxes. Commercial,
  off-the-shelf components are marked in orange. The dashed boxes surrounding
  components groups correspond to actual physical locations: the bigger
  ``HESS1U Camera`` box corresponds to the camera itself, the ``telescope
  shelter'' is a shelter in which the camera is parked during the day, the
  ``telescope hut'' is a white container mounted on the structure of the
  telescope, the ``farm'' is a server room located in the central control
  building of the array.}
  \label{fig:architecture}
\end{figure}

Single devices all have $100\eh{Mbit/s}$ copper-based Ethernet connections to a central switch,
which is then connected to the main camera server by means of a $10\eh{Gbit/s}$
optical fiber connection. As a novelty, the camera server is not located inside
the camera as it was in the past, but in the computer ``farm'' (see \fref{fig:architecture}), which eases its maintenance and prevents failures due to environmental factors.

\section{Software}

The presence of an ARM-based computer module running Linux on most of the camera systems makes communication
between single components easy to establish. In order to reduce the software development time, several open source software
solutions were adopted.

The operating system running on the Stamp9G45 is a version of Linux, build
using the Yocto embedded Linux build system \cite{yoctoproject}, with some
kernel patches provided by the manufacturer.
Slow control communication is implemented using Apache Thrift, a
lightweight and performant RPC framework \cite{apachethrift}.
Data transfer was implemented using the \zmq \ \cite{zmqsite} smart socket
message-passing library as a transport backbone. Raw event data messages are
serialized via an optimized custom protocol, and monitoring data messages are
serialized using the Google Protocol Buffers library \cite{protobuf}. A big advantage of the above RPC and serialization libraries is the automatic code generation they provide, which shortens the development cycles and is very stable against human programming errors.

The software in the various Linux systems of the camera is kept synchronised and up-to-date by installing one central instance of it on the camera server, and mounting it remotely in the file system of each device after start-up. This software also includes a test suite for each device that can be run any time as a self-check of basic functionality.

\section{Summary and Outlook}

Four new and NECTAr-based Cherenkov cameras for the 12-meter telescopes of \hess\ have been deployed in 2015/16, substantially reducing the dead time of the cameras and allowing for a more versatile and efficient operation and maintenance of the cameras. An example event recorded by the cameras is shown in \fref{fig:event}, left.

Besides providing the experiment with a newer technology and a much reduced readout dead time, the new cameras allow for more sophisticated and flexible algorithms than previously to process the trigger and gamma-ray signals in the front-end part of the camera.

A second contribution to this conference reports on the performance of the new cameras presented here \cite{hess1u_perf}.


\begin{thebibliography}{99}
\bibitem{hess_crab} Aharonian, F., et al., ``{Observations of the Crab nebula with HESS}'', {\em A\&A}~{\bf 457}~{899} (2006)

\bibitem{oldcameras} P.~Vincent et~al.,  ``{Performance of the H.E.S.S. cameras}'', {\em Proc. 28th ICRC} (2003)

\bibitem{hess1u_perf} Bonnefoy, S., et al. ``{Performance of the upgraded H.E.S.S. cameras}'', \pos{PoS(ICRC2017)805}

\bibitem{nectar}
Naumann, C.~L. et~al., ``{New electronics for the Cherenkov Telescope Array
  (NECTAr)}'', {\em NIM A}~{\bf 695},  44 (2012).
  
\bibitem{yoctoproject} ``{Yocto project -- Open Source embedded Linux build system, package metadata and SDK generator}'', \href{}{https://www.yoctoproject.org/}, Accessed: 2016-05-28  

\bibitem{apachethrift} ``{Apache Thrift - Home}'', \href{}{https://thrift.apache.org/}, Accessed: 2016-05-28 

\bibitem{zmqsite} ``{Code Connected - zeromq}'', \href{}{https://zeromq.org/}, Accessed: 2016-05-28 

\bibitem{protobuf} ``{Protocol Buffers}'', \href{}{https://developers.google.com/protocol-buffers}, Accessed: 2016-05-28 




\end{thebibliography}
\end{document}